**Local Multimodal Dynamics in Mixed Ionic-Electronic Conductors and Their Fingerprints in Organic Electrochemical Transistor Operation**

Shubham Tanwar[1,†,*], Han-Yan Wu[2], Chi-Yuan Yang[2], Ruben Millan-Solsona[1,3,‡], Simone Fabiano[2], Adrica Kyndiah[4,*], Gabriel Gomila[1,3]

[1]Nanoscale Bioelectrical Characterization Group, Institut de Bioenginyeria de Catalunya (IBEC), The Barcelona Institute of Science and Technology (BIST), Carrer Baldiri i Reixac 11-15, Barcelona, 08028 Spain

[†]Present address: Center for Nano Science and Technology, Istituto Italiano di Tecnologia, Via Rubattino 81, Milano, 20134 Italy (*email: shubham.tanwar@iit.it)

[2]Laboratory of Organic Electronics, Department of Science and Technology (ITN), Linköping University, Norrköping, SE-60174 Sweden

[3]Department d'Enginyeria Electrònica i Biomèdica, Universitat de Barcelona, Carrer Martí i Franquès, 1, Barcelona, 08028 Spain

[‡]Present address: Center for Nanophase Materials Sciences, Oak Ridge National Laboratory, Oak Ridge, Tennessee 37831, United States

[4]Center for Nano Science and Technology, Istituto Italiano di Tecnologia, Via Rubattino 81, Milano, 20134 Italy (*email: adrica.kyndiah@iit.it)

Correspondence should be addressed to: **Shubham Tanwar** and **Adrica Kyndiah**

**Mixed ionic-electronic conductors host tightly coupled interactions among mobile ions, electronic charges, and the polymer matrix, giving rise to complex multimodal responses spanning electrical, mechanical, and morphological transformations. These materials underpin organic electrochemical transistors (OECTs), which translate such interactions into low-voltage signal amplification and sensing for applications in bioelectronics, neuromorphic computing, and memory. Despite their central role, OECT current-voltage transfer characteristics are often treated phenomenologically, as both the local multimodal dynamics and their connection to global device response remain unresolved. Here, we reveal that the transfer curve encodes a cascade of spatially localized electrochemical transitions, each associated with distinct changes in conductivity, stiffness, and morphology, fundamentally redefining it as a spatially resolved fingerprint of device's internal state. Using automated operando multimodal in-liquid scanning dielectric microscopy, we directly map these dynamics and identify region-specific electrochemical thresholds governing the interplay between source, channel, and drain. We found that the local tip-sample electrostatic force serves as a remarkable mechanistic observable of coupled multimodal dynamics in mixed conductors. A physically grounded model links it to general material, interfacial, and geometric parameters, enabling mechanistic interpretation and predictive insights. Our work provides a new framework for probing and understanding mixed conduction in ion-electron coupled systems.**

Deciphering how local physicochemical processes give rise to macroscopic device response remains a central challenge in ion-electron coupled systems across diverse material platforms.[1–4] This challenge is particularly evident in multifunctional materials such as organic mixed ionic-electronic conductors[2] (OMIECs), where mobile ions, electronic carriers, and the soft polymer matrix interact in a tightly coupled manner.[5–8] Electrochemical ion insertion drives electronic charge compensation, altering local conductivity,[9] while simultaneously modulating mechanical properties through polymer softening or stiffening,[10] and inducing morphological reorganization via mass transport and local electrostatic interactions.[11] These multimodal responses underpin applications ranging from bioelectronics[12] to neuromorphic computing,[13] energy storage,[14] and soft robotics,[15] but they also complicate efforts to establish a unified mechanistic framework.[6–8,16] While individual electrochemical transformations in OMIEC thin films have been extensively characterized using optical,[17–19] X-ray,[20–23] and scanning probe techniques,[24–28] such approaches typically probe single modalities, and for multimodal dynamics, one has to rely on comparisons across disparate sample architectures, environments, and spatial resolutions.[6] Such fragmentation disrupts the coupling of processes, obscuring how they act together. Therefore, capturing and elucidating the intrinsic multimodal dynamics demands a probe capable of simultaneously resolving electrical, mechanical, and morphological changes while preserving their natural interplay and disentangling their distinct contributions.[6]

Because local multimodal dynamics of OMIEC materials remain unresolved, their connection to macroscopic device function is also poorly understood.[5] Organic electrochemical transistors[29] (OECTs) provide a natural platform to study this link, as their operation directly reflects the electrochemical control and readout of OMIEC behavior, making them a key building block of emerging technologies.[12,13,30,31] The OECT response is most commonly described by its current-voltage (I-V) transfer characteristics, which track the source-drain current as a function of gate bias at a constant drain voltage.[16] Unlike conventional field-effect transistors, where the I-V response is primarily governed by surface electrostatics,[32,33] OECT response arises from emergent behavior involving complex multimodal interactions described previously. Yet, transfer curves remain largely phenomenological, typically used to derive benchmark metrics[34] rather than as a window into the underlying local doping dynamics.[5] This limitation is also based on the inability to directly translate insights inferred from uniform thin-film studies to real device architectures, where electric fields and ionic fluxes vary spatially, causing region-specific changes. Recent machine-learning approaches have begun to exploit the embedded complexity in the full transfer curve shape, rather than relying solely on derived metrics, to enhance sensing performance in related electrolyte-gated devices,[35] underscoring that conventional models overlook critical aspects of device behavior.[36,37] While these approaches are practically powerful, they provide limited mechanistic insight. A comprehensive understanding, therefore, requires a framework that directly resolves local multimodal dynamics and connects them to the macroscopic I-V characteristics of operating devices.

Here, we address these challenges by establishing an experiment-interpretation framework that probes both local and global responses. Using automated multimodal in-liquid scanning dielectric microscopy (SDM),[33,38,39] we directly map spatially resolved electrical, mechanical, and morphological dynamics in operating OECTs under varying gate and drain biases. We use poly(benzimidazobenzophenanthroline) (BBL) polymer[40] as a model mixed conductor, where compositional homogeneity enables direct access to the intrinsic multimodal response. Moreover, BBL is among the few polymers that can sustain exceptionally high doping levels,[41] thereby also enabling the investigation of phenomena associated with high charge-carrier densities. The resulting data reveal distinct voltage dependencies across modalities, demonstrating that device function emerges from co-evolving spatially distributed dynamics (**Figure 1**), which renders conventional single-modality interpretations insufficient. A physically grounded equivalent circuit model of multimodal in-liquid SDM links these local patterns to characteristic conductivity transitions across source, channel, and drain regions, which in turn map onto key features of the transfer curve such as extrema in its first and second derivatives (**Figure 2**). This approach not only imparts mechanistic meaning to the I-V transfer curve shape but also uncovers unexpected ion-dependent behaviors. For example, $NH_4^+$ doping reduces the effective interfacial capacitance relative to even ion-poor Milli-Q water, arising from suppressed swelling at high doping levels (**Figure 3**). By reframing the transfer curve into a mechanistic map of local multimodal dynamics, our results provide a blueprint for connecting local physicochemical processes to global device behaviour, guiding the rational analysis and design of multifunctional ion-electron coupled systems.

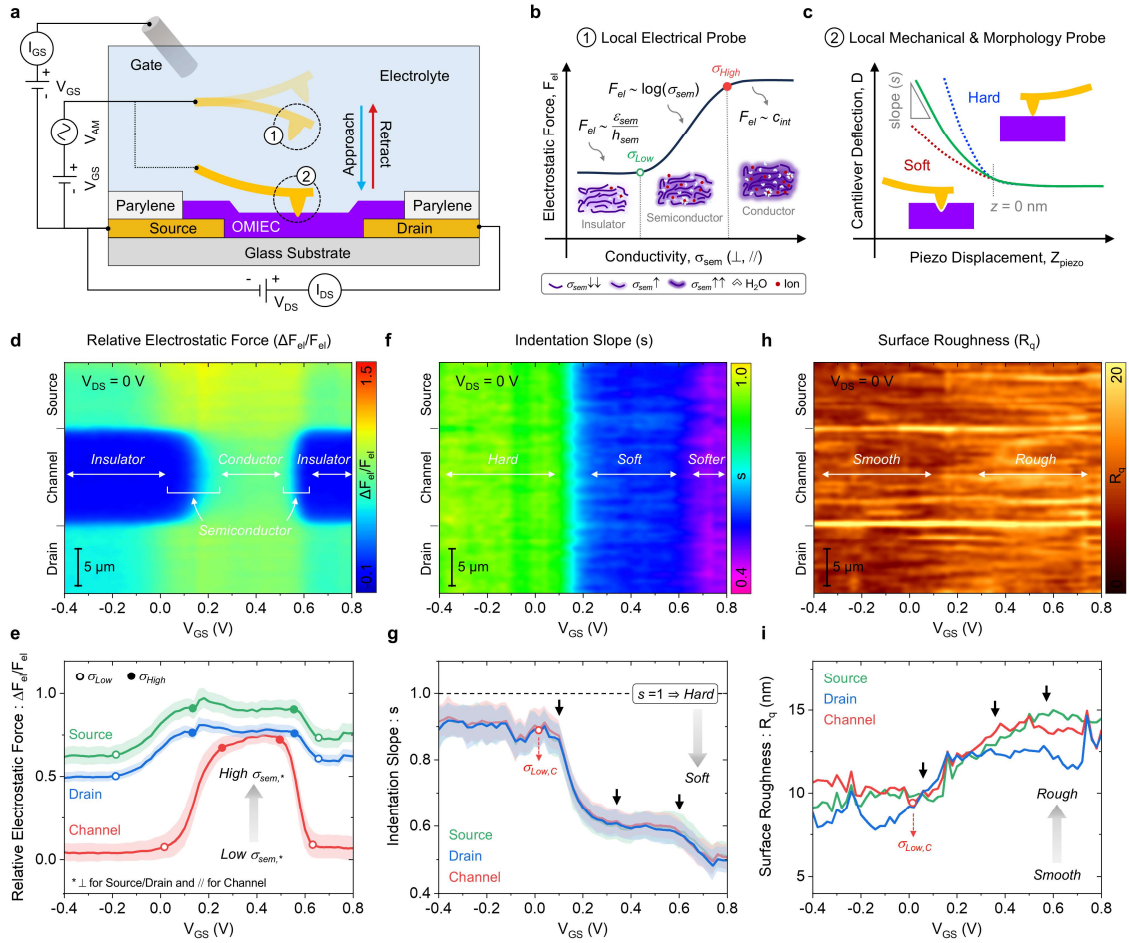

**Figure 1. Multimodal probing reveals bias-induced ionic-electronic dynamics. a-c**, Schematic of multimodal in-liquid scanning dielectric microscopy (SDM) measurements on an operating OECT (channel length $L = 10$ μm, width $W = 100$ μm) in 10 mM NaCl aqueous electrolyte (**a**). A gold-coated AFM tip serves as both the SDM probe and gate electrode, applying $V_{GS}$ together with an external tungsten electrode. During approach-retract cycles, the tip simultaneously probes local electrostatic force (**b**, sensitive to local conductivity), mechanical stiffness, and topography (**c**, reflecting polymer softening and swelling), capturing the local physical fingerprint of coupled ionic-electronic dynamics. **d-i**, Gate-dependent spatially resolved spectra and region-averaged trends of electrostatic force (**d,e**), mechanical indentation slope (**f,g**), and surface roughness (**h,i**) at zero drain bias ($V_{DS} = 0$ V). Region-averaged trends (**e, g, i**) are shown for source (green), channel (red), and drain (blue) regions, with shaded bands indicating spatial variability (standard deviation). In **e**, empty and filled circles mark local insulator-to-semiconductor ($\sigma_{Low}$) and semiconductor-to-conductor ($\sigma_{High}$) transitions, linked to transversal conduction in source/drain and lateral conduction in channel within the OMIEC thin-film. Arrows in **h** and **i** denote bias-driven changes in mechanical

stiffness (hard to soft) and surface morphology (smooth to rough). Together, panels **d-i** illustrate how multimodal mapping resolves the coupled electrical, mechanical, and morphological responses under electrochemical gating. Additional experimental parameters: electrostatic force generated using a carrier signal of $f_{el}$ = 115 MHz (modulated at $f_{mod}$ = 10 kHz, $V_{ac}$ = 1 VPP). The relative electrostatic force variation ($\Delta F_{el}/F_{el}$) is plotted for a tip-sample lift distance $Z_{lift}$ = 100 nm with reference distance $Z_{ref}$ = 1800 nm. Spatially resolved images (64 × 12 pixels, covering 30 × 5.625 µm$^2$) were acquired at each gate bias and used to reconstruct the spectra. Measured cantilever properties: spring constant $k$ = 1.20 N/m, deflection sensitivity $s_0$ = 24.19 nm/V, and resonance frequency $f_{res}$ = 36.45 kHz. See *Methods* for details on the experimental setup, signal origins, and data analysis procedures.

**Multimodal probing reveals bias-induced ionic-electronic dynamics**

Mixed ionic-electronic conductors rely on spatially distributed doping processes, but how they unfold in working devices across physical observables remains unresolved. To investigate this phenomenon across the nanoscale to microscale dimensions of the device, we utilized state-of-the-art n-type accumulation mode BBL OECTs with a conventional device architecture (channel length $L$ = 10 µm, width $W$ = 20 µm or 100 µm), as shown in optical and AFM images in **Figure S1a**. The need for multimodal characterization becomes experimentally evident when observing the pronounced electrochemically induced swelling of the BBL polymer. Direct thickness assessment with AFM reveals that the polymer expands from ~25 nm in the dry state to ~36 nm upon passive hydration in 10 mM NaCl electrolyte, and further swells up to ~80 nm under applied gate bias (**Figure S1b,c**). This substantial volume change of over a threefold increase in thickness highlights that morphological evolution and associated mechanical softening due to ion and electrolyte ingress are fundamentally linked to the electrochemical doping process and must be probed alongside electrical behavior.[10]

To probe these collective dynamics during device operation, we developed an automated[38] operando multimodal scanning probe technique (schematically illustrated in **Figure 1a** and detailed in the *Methods* section "*Multimodal operando in-liquid scanning dielectric* *microscopy*") that enables simultaneous nanoscale mapping of local electrical properties via electrostatic force (**Figure 1b**), mechanical properties via cantilever indentation (**Figure 1c**), and surface topography via cantilever deflection in an aqueous electrolyte. The electrostatic force signal is primarily sensitive to the local dielectric response, which is strongly influenced by the semiconductor conductivity at the measurement frequency,[33,39] in a manner analogous to AC impedance measurements, but with much higher spatial resolution and sensitivity (Figure 1b). The cantilever indentation probes local polymer softening during ion uptake (Figure 1c), and the cantilever deflection tracks corresponding morphological changes linked to local swelling and microstructural reorganization. The specific manifestations of these signal tendencies are examined in subsequent sections through comprehensive multimodal measurements and modelling. Together, these mappings provide a direct physical fingerprint of the doping process during electrochemical gating. We performed initial measurements on a pristine device under controlled gate bias ($V_{GS}$ was varied from -0.4 V to +0.8 V in 20 mV steps, comprising 61 bias points, to span low to extreme doping levels) with zero drain voltage ($V_{DS}$ = 0 V). At each fixed bias point, multimodal maps were acquired from a representative device region and subsequently compiled into a spectrum to track their evolution with $V_{GS}$ (see **Figure S1d** for the data compilation process). Typical cantilever approach curves acquired during this imaging sequence are shown in **Figure S2**, illustrating how variations in $V_{GS}$ influence the curves in a non-trivial manner. The initial use of zero drain bias eliminates lateral electric field gradients, enabling a direct view of how gate-induced doping evolves transversely across the source-channel-drain axis.

As gate voltage increases to more positive $V_{GS}$, we observe a sequence of distinct regimes, each reflecting complex bias-dependent evolution across electrostatic (**Figure 1d,e**), mechanical (**Figure 1f,g**), and morphological signals (**Figure 1h,i**). Initially, the polymer is electrostatically inert, mechanically rigid, and morphologically smooth, consistent with an undoped insulating state. Beyond a local threshold voltage (corresponding to a conductivity denoted as $\sigma_{Low}$ in Figure 1e), the electrostatic force increases sharply, indicating the onset of bulk conductivity as mobile charge carriers begin to screen the AC probe field. The $\sigma_{Low}$ transition marks the emergence of a partially doped semiconducting state. As the polymer gets electrochemically doped, it mechanically begins to soften (Figure 1g) and topographically roughen (Figure 1i), revealing active ion ($Na^+ \cdot H_2O$) uptake and material reorganization. At intermediate gate voltages (between $\sigma_{High}$ points in Figure 1e), a mechanically stable and electrically conductive state develops, exhibiting a plateau where ionic-electronic dynamics appear to be equilibrated. However, at higher gate voltages ($V_{GS} > 0.6$ V), all three modalities evolve again: the electrostatic force (i.e., conductivity) decreases, while mechanical softening and roughness continue. The observed decrease in conductivity at higher $V_{GS}$ suggests the onset of an over-doped state, where increasing morphological disorder and Coulombic interactions between charge carriers reduce their mobility.[42] Such a conductivity collapse at high doping levels, previously inferred from a decrease in $I_{DS}$ with increasing $V_{GS}$ and described as *anti-ambipolar* behavior,[41] is now directly resolved through spatially mapped multiphysical processes using our multimodal approach.

To interpret these measurements, we developed a physically grounded equivalent circuit model of the tip-electrolyte-device system (see **Figure S3** and *Methods* section "*Equivalent circuit model of in-liquid SDM for EGTs*" for details). The model relates the measured electrostatic force to key material parameters, including local semiconductor conductivity ($\sigma_{sem}$), effective interfacial capacitance ($c_{int}$) reflecting bulk ionic-electronic coupling, semiconductor dielectric constant ($\varepsilon_{sem}$), and electrochemical swelling ($h_{sem}$), thereby clarifying how distinct doping regimes emerge (see Figure 1b for a visual summary of their influence). It identifies the conditions where out-of-plane dielectric response, in-plane conduction, and ionic-electronic screening dominate the tip-sample interaction. Because the model is formulated entirely in terms of general material, interfacial, and geometric parameters, it provides a broadly applicable interpretation framework for understanding local responses in mixed ionic-electronic conductors.

From the perspective of the SDM probe, the model identifies two characteristic local conductivity thresholds ($\sigma_{Low}$ and $\sigma_{High}$ in Figure 1b) in each device region, which can be intuitively understood as local *insulator-to-semiconductor* and *semiconductor-to-conductor* transitions (see Figure 1b). We collectively denote these thresholds as $\boldsymbol{\sigma_T} = (\sigma_{Low,\boldsymbol{T}}, \sigma_{High,\boldsymbol{T}})$, where $\boldsymbol{T} \in \{S, C, D\}$ corresponds to the source, channel, and drain regions, respectively; the same notation is used for the corresponding gate voltages at which these transitions occur. Experimentally, these thresholds appear as knee points in the electrostatic force response (Figure 1e): $\sigma_{Low}$ marks the onset of sensitivity to local conductivity, while $\sigma_{High}$ indicates the onset of saturation. As detailed in the *Methods*, this sigmoidal response naturally arises from conductivity-controlled electrodynamic processes, specifically dielectric relaxation, Maxwell-Wagner polarization, and voltage partitioning within the tip-sample system, giving rise to three conductivity-defined regimes (dielectric, lossy semiconducting, and effectively metallic). While the numerical values of the conductivity thresholds depend on the material, the underlying regime structure and resulting sigmoidal behavior are general and, in principle, apply to other mixed conductors, including p-type and ambipolar systems, though experimental validation in such systems is not included here.

In the present BBL-based OECT, the full measured response exhibits an inverted U-shape, consisting of two such consecutive sigmoidal segments: an initial rise (-0.4 V ≤ $V_{GS}$ ≤ +0.4 V in Figure 1e) followed by a decline at higher gate voltages (+0.4 V ≤ $V_{GS}$ ≤ +0.8 V in Figure 1e). The presence of the second sigmoidal segment reflects the ability of BBL to sustain high doping levels, allowing us to probe high-carrier-density regimes that would not be accessible in materials with limited electrochemical stability. These two segments exhibit similar electrical conductivities but display mechanically and morphologically distinct low- and high-doping regimes, as observed in Figures 1g and 1i. The exact physical nature of the $\sigma_T$ transitions is discussed in detail in the *Methods*. Importantly, these transitions are anisotropic, reflecting the orientation of the probe's AC electric fields relative to the sample. Over the source and drain regions, the fields couple primarily in the vertical direction, so the observed transitions reflect changes in *out-of-plane* (denoted as "⊥") conduction. In contrast, within the channel, the insulating substrate redirects the fields laterally, allowing the transitions to probe *in-plane* (denoted as "//") conduction along the channel axis.[33,43] This spatial anisotropy in transition sensitivity mirrors the direction of current flow from source to drain under operating bias, enabling the locally defined $\sigma_T$ transitions to map directly onto the macroscopic I-V transfer characteristics.

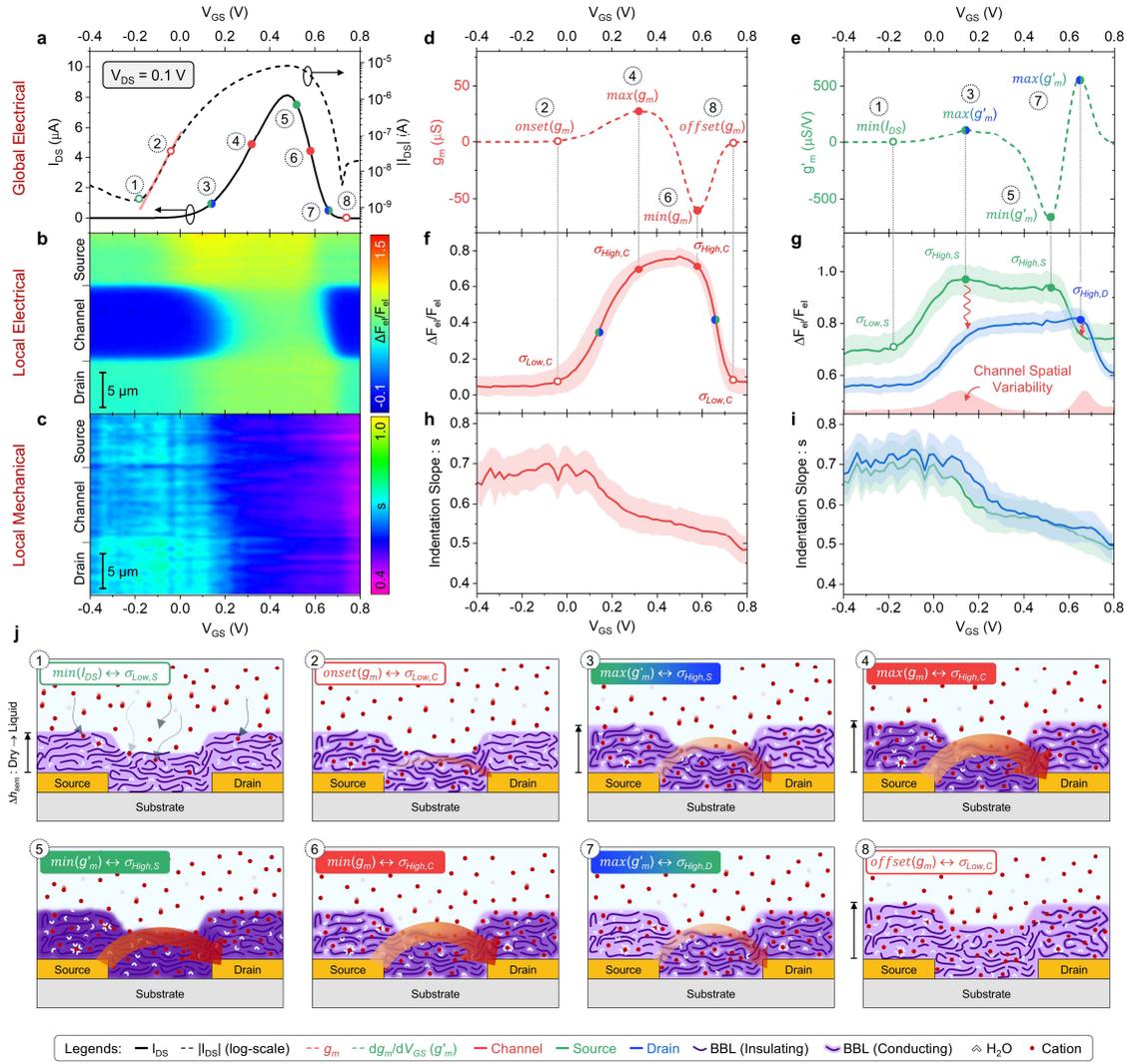

**Figure 2. Fingerprints of spatially resolved doping dynamics in OECT transfer characteristics. a**, Drain current ($I_{DS}$) versus gate voltage ($V_{GS}$) transfer curve at $V_{DS}$ = 0.1 V shown in linear (continuous black line) and semi-log (dashed black line) scales, reconstructed from $I_{DS}$ vs *time* traces acquired during multimodal mapping. **b,c**, Associated gate-dependent spatially resolved spectra of electrostatic force (**b**) and mechanical indentation slope (**c**). **d,e**, First derivative (**d**, transconductance $g_m$) and second derivative (**e**, d$g_m$/d$V_{GS}$ or $g'_m$) of the transfer curve in **a**, with key operating points (minima/maxima) linked by vertical dashed lines to corresponding local physical states ($\sigma$ points). **f-i**, Region-averaged trends of electrostatic force (**f**, **g**) and indentation slope (**h**, **i**) for channel (red), source (green), and drain (blue) regions. Shaded bands indicate spatial variability (standard deviation), with channel variability overlaid in **g** to highlight correlations with positive curvature points in **e**. In **g**, wiggly arrows mark points where channel control shifts between source and drain. **j**, Schematic representation of local physical states at eight key operating points marked in **a**, linking global electrical

response to spatial doping dynamics: (1) onset of transversal conduction ($\sigma_{sem,\perp}$) and exponential carrier rise ($I_{DS}$ minimum), (2) channel formation (red arrow) initiating $g_m$ ($\sigma_{sem,//}$) rise and marking device turn-on voltage, (3) peak $g'_m$ curvature linked to maximum spatial heterogeneity with source and drain competing for channel control, (4) fully formed channel yielding peak $g_m$, (5) peak negative $g'_m$ curvature linked to mobility loss at source due to high-doping causing effective source-drain role reversal, (6) negative peak $g_m$ mirroring (4) with role reversal, (7) peak $g'_m$ mirroring (3) with role reversal, (8) overdoping collapses channel conductivity and transitioning it into a soft insulating state. Additional experimental parameters are as in Figure 1.

**Fingerprints of spatially resolved doping dynamics in OECT transfer characteristics**

Introducing a non-zero $V_{DS}$ leads to a spatially varying lateral electric field in the device,[33] which modulates the local effective gate potential and alters doping dynamics along the source-channel-drain axis. The lateral electric field drives mobile charge carriers from source to drain, resulting in a source-drain current ($I_{DS}$) that varies with $V_{GS}$, yielding the characteristic transfer curve (**Figure 2a** for $V_{DS}$ = 0.1 V). This transfer curve is reconstructed from $I_{DS}$ vs *time* traces acquired during multimodal mapping at varying $V_{GS}$ and fixed $V_{DS}$ (see **Figure S4**). The drain current in these $I_{DS}$ vs *time* traces remains relatively stable during multimodal imaging, confirming that the SDM probe does not significantly perturb global device behavior. The associated multimodal measurements reveal that $V_{DS}$ breaks the device's initially homogeneous response and induces a spatial variability in both electrostatic (**Figure 2b**) and mechanical (**Figure 2c**) response. It reflects the non-uniform distributions of mobile electronic carriers and ionic species, respectively, which become more pronounced with increasing $V_{DS}$ (see **Figure S5** for measurements at different $V_{DS}$ values, ranging from 0.2 V to 0.5 V in steps of 0.1 V). The observed spatial modulation provides direct evidence that the coupled evolution of electronic and ionic charge distributions shapes the *electrochemical potential* landscape of an operating device.[37] Unlike electrostatic potential, the electrochemical potential captures both electrical and chemical contributions, making it the primary driving force behind charge transport and local transitions in OECTs.

From local mapping, we can readily identify region-specific characteristic $V_{GS}$ for conductivity transitions $\sigma_T$ defined previously. Knowing these region-specific *thresholds* allows us to determine which parts of the device are insulating, semiconducting, or conducting, and thereby identify the regions that dominate the overall device response at any given operating point. A natural question follows: can these region-specific $V_{GS}$ values be inferred from the transfer curve alone? If so, they will establish a route for the mechanistic interpretation of local device dynamics directly from global I-V characteristics, grounded in insights uniquely revealed by spatially resolved measurements. This local-global correspondence is particularly relevant because current physical models of OMIECs do not yet capture all relevant modalities and effects, including transport anisotropy that couples vertical conduction over the source and drain to horizontal conduction along the channel, as well as their interactions across the full device architecture as discussed here. While our framework fully explains the local behavior in the source, channel, and drain regions, a comprehensive model that integrates these regions and predicts the complete transfer curve shape is not yet available. In the absence of a comprehensive local-to-global theoretical framework, we aim to empirically identify patterns in how bias-dependent, spatially resolved internal states (see **Figure S6**) evolve and collectively shape the global transfer characteristics, providing valuable insights into the operational multiscale mechanism of OECTs.

Mathematically, the transfer curve shape is characterised through its slope (transconductance $g_m$, **Figure 2d**) and curvature (derivative of transconductance $dg_m/dV_{GS}$ or $g'_m$, **Figure 2e**). Transconductance indicates how sensitively the channel current responds to gate modulation, while its derivative highlights where this sensitivity accelerates or declines. We observed that extrema points in these quantities align closely and uniquely with region-specific $\sigma_T$ conductivity transitions identified in the local electrostatic force response (**Figure 2f,g**), as depicted by the vertical dashed lines in Figure 2d-g, suggesting a direct physical link between local state transitions and transfer curve evolution. These correlations consistently appear across measurements (see Figure S5).

The initial onset of $g_m$, where noticeable longitudinal channel conductivity ($\sigma_{sem,//}$) first permits considerable $I_{DS}$ flow, occurs when the channel surpasses its $\sigma_{Low,C}$ threshold, establishing a physically grounded definition of device turn-on or threshold voltage. The maximum $g_m$ aligns with the transition to $\sigma_{High,C}$, when the longitudinal channel is fully formed (see vertical dashed lines in Figure 2d,f). The shape of the transfer curve further reflects the interplay between source and drain regions in governing channel formation (see vertical dashed lines in Figure 2e,g). The $\sigma_{Low,S}$ state of the source corresponds to the $I_{DS}$ minimum, marking the initial rise of transversal conductivity ($\sigma_{sem,\perp}$) in the OMIEC thin-film. Points of positive curvature ($g'_m$) exhibit mixed character, reflecting the competition between source and drain in setting the electrochemical potential distribution along the channel. As a result, the channel shows the highest spatial variability at these curvature maxima (see overlaid standard deviation of the channel electrostatic force in Figure 2g). Two such maxima are observed: one in the low-doping regime, associated with $\sigma_{High,S}$ at the source, where increasing $V_{GS}$ gradually shifts control of the channel toward the drain (indicated by a wiggly arrow in Figure 2g); and another in the high-doping regime, associated with $\sigma_{High,D}$ at the drain, where increasing $V_{GS}$ progressively shifts control toward the source (also indicated by a wiggly arrow in Figure 2g). In contrast, the negative curvature point corresponds to the $\sigma_{High,S}$ state of the source in the high-doping regime. This correlation reveals that anti-ambipolar characteristics, where $I_{DS}$ first rises and then falls with increasing $V_{GS}$, which leads to negative curvature, originate from asymmetric local conductivity collapse. At high gate voltages, over-doping of the source reduces charge carrier mobility through Coulombic interactions,[42] thereby decreasing the electrostatic force and shifting the conduction pathway towards the drain. This dynamics effectively reverses the source and drain roles, rendering the drain more conductive than the source and thereby mimicking p-type transport in an otherwise n-type device.[42] Minima in $g'_m$ capture this role reversal, demonstrating that the curvature of the transfer curve encodes direct physical meaning. Overall, as indicated in Figure 2a, the patterned alignment of *insulator-to-semiconductor* (empty colored circle symbols) and *semiconductor-to-conductor* (filled colored circle symbols) transitions across the source, channel, and drain reveals an emergent global structure in otherwise local electrochemical dynamics. This ordered overlap along the transfer curve demonstrates that macroscopic device operation arises from coordinated region-specific transitions rather than spatially averaged behavior.

Simultaneously acquired mechanical measurements complement this interpretation. Ion uptake softens the polymer progressively, reaching saturation near the $g_m$ peak and continuing to evolve as over-doping sets in (**Figure 2h**). Over source and drain, the mechanical signal onset occurs at different $V_{GS}$ due to the applied $V_{DS} = 0.1$ V bias (**Figure 2i**). At $V_{GS} > 0.7$ V, total conductivity collapse coincides with pronounced polymer softening (purple-colored region in Figure 2c), reconfirming that morphological disorder contributes to the disruption of charge transport. However, this increased softening at high doping sensitively depends on the ionic species involved, as shown by the absence of additional swelling with $NH_4^+$ doping investigated in the next section.

Beyond these gradual trends, we observe sharp and spatially extended abrupt changes near the $\sigma_{High,C}$ points at high $V_{DS}$, coinciding with the $g_m$ peak (for instance, see the $V_{DS} = 0.3$ V case in Figure S5b). This abrupt change resembles a phase-like transition in the polymer's physical state, as evidenced by both its electrical and mechanical responses. This phase-like transition may arise from emergent collective effects, possibly including ion crowding and steric constraints,[44] and correlated ionic-electronic rearrangements at the electrolyte-polymer interface, which become energetically favorable only once a critical local field or charge density is reached, such as at peak $g_m$ conditions.[45] While the precise origin and microscopic mechanism remain to be clarified, our observations indicate that the characteristic $\sigma_T$ thresholds act as critical points at which the material undergoes a rapid reorganization to relieve accumulated electrochemical stress.

Together, these spatial fingerprints establish a mechanistic framework in which the OECT transfer curve (Figure 2a) serves as a unique map of local electrochemical transitions, enabling a physically grounded understanding of device operation, as schematically summarised in **Figure 2j**. In parallel, such local measurements also enable the interpretation of distinct diode-like[46] output characteristics ($I_{DS}$ vs $V_{DS}$ curve at fixed $V_{GS}$) of BBL OECTs, which shift to higher $V_{DS}$ with increasing $V_{GS}$ in the high-doping regime (**Figure S7**). We next analyse ion-specific effects on doping dynamics through this mechanistic interpretation of transfer curves, and simultaneously validate them through local multimodal measurements.

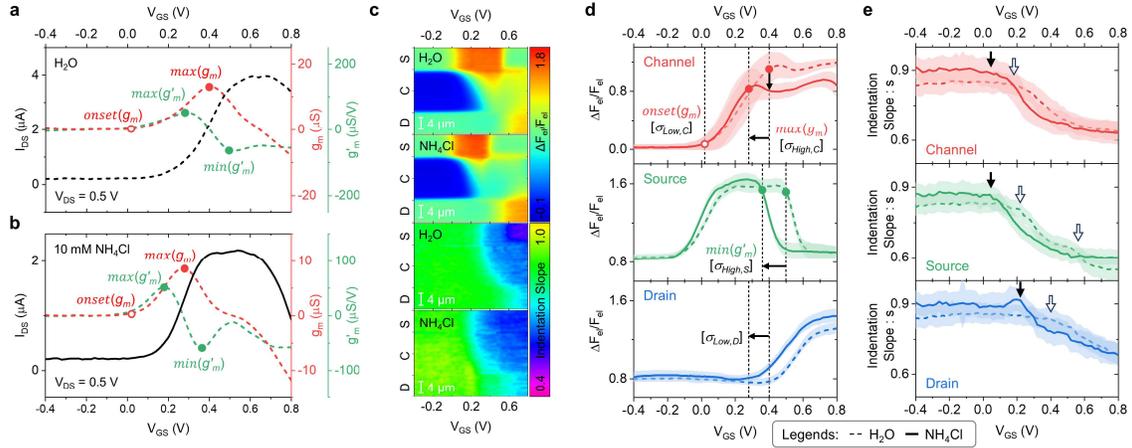

**Figure 3**. **Ion-specific modulation of doping dynamics revealed through transfer curve shape. a,b**, Reconstructed transfer curves of an OECT (channel length $L$ = 10 μm, width $W$ = 20 μm) operating in Milli-Q water (**a**, dashed black line) and 10 mM NH$_4$Cl electrolyte (**b**, continuous black line) at $V_{DS}$ = 0.5 V, with first ($g_m$, red dashed line) and second ($g'_m$, green dashed line) derivatives. Shifts in minima and maxima of $g_m$ and $g'_m$ (correlated to local $\sigma_T$ points) enable comparison of doping dynamics across electrolytes. Earlier onset of $g'_m$ extrema ($\sigma_{High,S}$) in NH$_4$Cl indicates faster polymer doping, while reduced $g_m$ peak ($\sigma_{High,C}$) suggests lower effective interfacial capacitance. **c-e**, Multimodal measurements corroborating the trends observed in **a** and **b**. **c**, Gate-dependent electrostatic force (top) and indentation slope (bottom) spectra for both electrolytes. **d,e**, Region-averaged trends for channel (red), source (green), and drain (blue) extracted from **c** for electrostatic force (**d**) and indentation slope (**e**). Shaded bands indicate spatial variability (standard deviation) within each region; arrows in **e** (empty: Milli-Q water, filled: NH$_4$Cl) highlight electrolyte-dependent shifts in mechanical response. Spectra and averages collectively show spatially resolved doping dynamics and their dependence on ionic environment. Additional experimental parameters: electrostatic force generated using a carrier signal of $f_{el}$ = 105 MHz (modulated at $f_{mod}$ = 10 kHz, $V_{ac}$ = 2 VPP). The relative electrostatic force variation ($\Delta F_{el}/F_{el}$) is plotted for a tip-sample lift distance $Z_{lift}$ = 100 nm with reference distance $Z_{ref}$ = 1800 nm. Spatially resolved images (60 × 10 pixels, covering 26 × 4.33 μm$^2$) were acquired at each gate bias and used to reconstruct the spectra. Measured cantilever properties: spring constant $k$ = 0.9 N/m, deflection sensitivity $s_0$ = 32.68 nm/V, and resonance frequency $f_{res}$ = 31.29 kHz.

**Ion-specific modulation of doping dynamics revealed through transfer curve shape**

The chemical identity of ions in the electrolyte plays a defining role in shaping the device's operation by dictating how and when the OMIEC transitions between $\sigma_T$ states. In BBL polymer, this chemical influence is especially pronounced when comparing device responses in electrolytes containing ammonium ($NH_4^+$) ions[41] versus ion-poor Milli-Q water (**Figure 3**). At first glance, both environments support conventional $I_{DS}$ modulation with $V_{GS}$ (**Figure 3a,b**). In both cases, the onset of $g_m$ occurs at the same $V_{GS}$, indicating similar conductivity thresholds for channel formation ($\sigma_{Low,C}$). However, despite an ion-rich environment in $NH_4Cl$, the maximum $I_{DS}$ is reduced by half compared to ion-poor Milli-Q, suggesting a more nuanced interplay between ion ingress and charge transport.

Interpreting the transfer curve through $\sigma_T$ transitions resolve this apparent contradiction. The narrowed gate voltage window between the onset ($\sigma_{Low,C}$) and peak $g_m$ ($\sigma_{High,C}$) in $NH_4Cl$ reflects the faster development of a fully formed channel with $V_{GS}$, as well as the earlier onset of mobility loss at the source in high-doping state ($g'_m$ minima = $\sigma_{High,S}$). Once the channel exceeds its high-conductivity threshold ($\sigma_{High,C}$ = peak $g_m$), the device enters a regime where the local response is governed mainly by the effective Stern interfacial capacitance representing bulk ionic-electronic coupling (see Figure 1b and Figure S3c,f for the influence of $c_{int}$ on electrostatic force response at high conductivity). The diminished peak current in $NH_4Cl$ thus indicates a reduction in this capacitance. On the other hand, ion-poor Milli-Q water represents a higher $c_{int}$, suggesting that the difference lies in the chemical identity of $NH_4^+$ ion and its interfacial behavior with the polymer.

Local multimodal measurements reinforce this interpretation (**Figure 3c-e**). The lower plateau in the channel electrostatic force at high $V_{GS}$ in $NH_4Cl$ confirms a reduced $c_{int}$ (Figure 3d, top panel). The narrowing of the inverted U-shaped electrostatic force response over the source (Figure 3d, middle panel), together with the shift in the drain response (Figure 3d, bottom panel), indicates more rapid conductivity state transitions with $V_{GS}$, consistent with earlier ion ingress (Figure 3e). Yet, unlike in NaCl or Milli-Q water, no additional mechanical softening is observed in the source region at high $V_{GS}$ in NH$_4$Cl, despite the source entering an electrostatically over-doped regime. In Milli-Q, a pronounced softening feature appears over the source at high gate bias, visible as the purple-colored region in the mechanical spectra (Figure 3c; see also the source green curve in Figure 3e for $V_{GS} > 0.5$ V), whereas this feature is absent in NH$_4$Cl. Over the channel and drain regions, the electrostatic force does not decrease at high $V_{GS}$ due to the applied drain bias, indicating that these regions do not enter an over-doped regime. The absence of a mechanical response over the source in NH$_4$Cl despite deep doping implies that the morphological disorder, which typically accompanies conductivity collapse, is suppressed. The NH$_4^+$ doping thus appears to bypass the mechanical over-doping transition that normally leads to volumetric expansion. This restricted volumetric expansion gives rise to reduced $c_{int}$ and points to a distinct mode of polymer-ion interaction. Moreover, the onset of mechanical transition does not always align with that of electrical transition, indicating that, while mechanistically linked, they are governed by separate thresholds and distinct physical constraints. Lastly, the multimodal measurements acquired during the backward $V_{GS}$ sweep indicate that additional polymer softening in Milli-Q water is reversible (**Figure S8**), suggesting that the BBL polymer recovers easily from the over-doped state, evidencing its robust stability.

Taken together, these signatures indicate a chemically specific role of NH$_4^+$ in reshaping both the electrical and mechanical landscapes of the device. Its capacity to form hydrogen bonds with carbonyl and nitrogen groups in the BBL polymer likely facilitates rapid ionic compensation of electronic charges as the $V_{GS}$ increases.[41] At the same time, the absence of mechanical softening at high doping and the reduced $c_{int}$ suggest that these interactions may constrain volumetric swelling, perhaps by inducing local rigidity within the polymer matrix.[10] These findings highlight a broader principle: electrolyte chemistry, in conjunction with polymer design, provides a powerful lever to program the functional response of organic mixed ionic-electronic conductors.

**Conclusion**

We have introduced a multimodal operando framework that directly reveals the internal physical dynamics of mixed ionic-electronic conductors during transistor operation, and connects these local processes to the global response of organic electrochemical transistors. Using BBL-based OECTs as a model system, we show that local conductivity transitions, mechanical responses, and morphological reorganization emerge concurrently under electrochemical doping, evolving in a spatially structured manner across the source, channel, and drain. These bias-dependent internal states define a dynamic electrochemical landscape, whose evolution is directly embedded in the macroscopic transfer characteristics.

Central to this approach is the measured local tip-sample electrostatic force, which acts as a mechanistic descriptor of intrinsic multimodal dynamics by sensing how ionic-electronic coupling, transport pathways, and dielectric screening evolve under operating conditions, while being simultaneously corroborated by mechanical and morphological responses. Through a physically grounded equivalent-circuit model, we demonstrate that this force quantitatively reflects key material and geometric parameters, including interfacial capacitance, transport anisotropy, film thickness, and dielectric response, and distinguishes between vertical transport over the source and drain and lateral transport along the channel. In this way, local electrostatic force serves as a unifying observable that integrates electrical, mechanical, and morphological state changes within the active material.

By correlating these spatially resolved internal states with global transfer characteristics, we establish a direct correspondence between local region-specific conductivity transitions and features in the transconductance and its voltage derivative. This effectively transforms the transfer curve into a voltage-resolved fingerprint of the device's internal physical state, even in the absence of a complete local-to-global device model. Once established, this local-to-global correspondence also provides predictive insight into how changes in local material, interfacial, or ionic conditions manifest in global device response.

While demonstrated here in BBL-based OECTs, the framework is general and applicable to other mixed conductors, where ion specificity, transport anisotropy, and interfacial effects may differ but can be interrogated using the same experimental and interpretation strategy. Overall, this work provides both experimental access to the multimodal internal dynamics in mixed conductors and a transparent framework for understanding how these dynamics collectively shape macroscopic device behavior, thereby moving beyond conventional phenomenological descriptions toward a physically informed view of device operation. This provides a foundation for rational analysis and design of ion-electron coupled systems.

**Methods**

*Multimodal operando in-liquid scanning dielectric microscopy*: Multimodal operando in-liquid scanning dielectric microscopy (SDM) was implemented by extending a previously established automated SDM platform designed for electrolyte-gated organic field-effect transistors (EGOFETs).[38] While prior work focused exclusively on local electrical measurements, the present approach integrates simultaneous electrical, nanomechanical, and morphological characterization, providing a unified multimodal framework tailored for organic electrochemical transistors (OECTs). Unlike EGOFETs, where gating is confined to interfacial charge modulation with negligible morphological impact,[33] OECTs rely on bulk ion penetration into the semiconductor, inducing strongly coupled electrical, mechanical, and morphological transformations during operation.

Operando measurements were performed on OECTs with standard device architecture (channel length $L$ = 10 µm, width $W$ = 20 µm or 100 µm) in aqueous electrolytes, without requiring any architectural modifications. A conductive atomic force microscopy (AFM) tip functioned both as the gate electrode (in conjunction with an external tungsten electrode) and as the scanning probe for multimodal data acquisition. Force-distance approach curves were recorded in force-volume mode across the representative source-channel-drain region of the OECT at each applied gate ($V_{GS}$) and drain ($V_{DS}$) bias (see Figure S2 for representative cantilever oscillation amplitude and deflection approach curves acquired at varying $V_{GS}$ and zero $V_{DS}$). Out-of-contact cantilever oscillations were used to extract the electrostatic force, contact-region indentation was used to assess nanomechanical stiffness (from the cantilever deflection vs. piezo displacement slope), and the static deflection in contact was used to derive topography. Data acquisition was fully automated using custom Python routines,[38] enabling high-throughput bias-resolved multimodal mapping.

Cantilever electrostatic force oscillations were generated by applying a high-frequency amplitude-modulated AC voltage ($V_{ac}$) to the conductive AFM tip. The carrier frequency ($f_{el}$, in MHz) was selected to be above the dielectric relaxation frequency of the electrolyte, ensuring sensitivity to local dielectric and conductive changes in the polymer while minimizing the response from mobile ions in the electrolyte. The tip oscillation amplitude at the modulation frequency ($f_{mod}$, in kHz) was extracted from the AFM photodiode signal with a lock-in amplifier, providing a signal proportional to local tip-sample electrostatic interactions.

AFM probes were calibrated using the standard contact-free thermal noise method under electrically floating conditions prior to the multimodal measurements. Cantilever spring constant ($k$), deflection sensitivity ($s_0$), and resonance frequency ($f_{res}$) were extracted and used to convert raw signals into quantitative mechanical and electrostatic force values. Further technical details, including bias conditions, scanning areas, lift heights, and acquisition parameters, are provided in respective figure captions.

*Instrumentation*: The system is built around a JPK Nanowizard 4 BioAFM (Bruker) integrated into a Nikon Eclipse inverted optical microscope. OECT was mounted on a custom AFM sample holder and biased via point probes (EverBeing EB-700). Source-gate ($V_{GS}$) and source-drain ($V_{DS}$) voltages were applied using a Keysight B2912A source-measure unit. For electrostatic force detection, a high-frequency amplitude-modulated AC voltage was applied to a gold-coated conductive AFM tip (HQ:NSC19/Cr-Au MikroMasch, nominal tip radius < 35 nm) using a Keysight 33622A RF waveform generator. The modulation signal was provided by Anfatec eLockIn 204/2 lock-in amplifier. The applied waveform parameters (AC amplitude and DC offset) were measured with a Keysight DSOX3024T oscilloscope. Multimodal imaging was performed in advanced Quantitative Imaging (QI) mode, acquiring approach curves at each pixel within a predefined grid. Instrument control and synchronization were handled by a custom Python package.[38]

*Electrostatic force data analysis and representation*: The local electrostatic force signal was extracted from the amplitude of the cantilever oscillation at the modulation frequency of the applied high-frequency AC voltage. These signals reflect in an analogous manner the local impedance of the sample, which varies strongly across operating regimes, from insulating (off-state) to conductive (on-state). Such transitions alter the tip-sample impedance and consequently the AC voltage delivered to the tip, introducing systematic artifacts in the measured electrostatic force. To remove these artifacts, we recently introduced a calibration-free normalization scheme for in-liquid SDM measurements.[38] Instead of absolute values, this approach uses the relative electrostatic force variation:

$$\frac{\Delta F_{el}}{F_{el}} = \frac{F_{el,Z_{lift}} - F_{el,Z_{ref}}}{F_{el,Z_{ref}}} \qquad (1)$$

where $Z_{lift}$ is the tip-sample distance chosen close to the surface to maximize measurement locality, and $Z_{ref}$ is a reference distance far from the sample used for normalization. This procedure eliminates high-frequency artifacts caused by variations in the AC tip-sample voltage, resulting in a signal that accurately reflects local electrical changes. Importantly, this normalization cancels out all scaling factors (cantilever spring constant, deflection sensitivity, lock-in gain, and tip-sample AC voltage amplitude) and can be applied directly to raw cantilever oscillation amplitudes.[38]

The data were processed using a custom Python workflow that extracts $\Delta F_{el}/F_{el}$ values at predefined tip-sample distances estimated from the cantilever static-deflection approach curves. These normalized values were used to generate spatial maps, or "lift images", of the relative electrostatic force at each gate and drain bias. Pixels were assigned to source, channel, or drain regions based on topography-guided segmentation. Regional averages were plotted against gate voltage, with error bars indicating regional standard deviations.

To visualize spatially resolved electrostatic force evolution across gate bias, images were averaged along the slow (transverse) scan axis and concatenated across voltages to create two-dimensional spectral maps (see Figure S2d). These spectra capture gate-dependent changes in local electrical properties along the device length.

While quantitative modeling of electrostatic force response can be pursued using numerical finite-element simulations,[33,38,39] here we developed a physically grounded equivalent circuit model (see "*Equivalent circuit model of in-liquid SDM for EGTs*" in *Methods*) to interpret electrostatic force variations in terms of characteristic conductivity thresholds. These thresholds correlate with key operating points in the device transfer curve, linking local electrical states to macroscopic device behavior and enabling a mechanistic understanding of ionic-electronic dynamics.

*Nanomechanical data analysis and representation*: Nanomechanical properties were extracted from the indentation region of force-distance curves acquired during probe-sample contact. The indentation slope,

$$s = \frac{dD}{dZ_{piezo}} \qquad (2)$$

where $D$ is calibrated cantilever deflection (nm) and $Z_{piezo}$ is the piezo displacement (nm), quantifies the mechanical softness of the tip-sample interface.

Vertical deflection was calibrated using the cantilever deflection sensitivity ($s_0$, in nm/V), measured by the thermal noise method under electrically floating conditions. Indentation slopes were obtained via linear fits to the contact portion of the approach curve. These indentation slopes vary with material stiffness: $s \to 1$ indicates an effectively rigid (non-deformable) material for the given cantilever, while s < 1 corresponds to a softer material.

The indentation slope is related to the effective sample stiffness through the mechanical coupling between the cantilever and the sample. Under the assumption of linear elastic deformation, the force-indentation relationship simplifies to:

$$s = \frac{\frac{1}{k_{cantilever}} \left|\frac{dF}{d\delta}\right|}{1 + \left(\frac{1}{k_{cantilever}} \left|\frac{dF}{d\delta}\right|\right)} \tag{3}$$

where $dF/d\delta$ is the local force-to-indentation response of the sample. For soft materials, $dF/d\delta \approx k_{sample}$, and the equation can be rearranged to estimate the effective sample stiffness:

$$k_{sample} = k_{cantilever} \left(\frac{s}{1-s}\right) \tag{4}$$

This provides a direct estimate of the effective sample spring constant from measurements (**Figure S9**). It is important to note that $k_{sample}$ reflects the effective stiffness of the entire device interface, including substrate contributions, rather than the OMIEC layer alone. This effective stiffness is functionally relevant for bioelectronics, as it determines how the device mechanically couples to biological tissue. Although substrate influence on nanomechanical measurements, commonly known as the bottom-contact effect, is well documented in AFM literature,[47] we do not correct for it here, as the measured stiffness accurately represents the mechanical environment experienced during device operation.

Regional trends and spectral maps for indentation slope were generated using the same Python pipeline and segmentation scheme described above for electrostatic force.

*Morphological data analysis and representation*: Surface topography was obtained from the static cantilever deflection at constant force recorded during contact, yielding height values at each pixel. Morphological changes were quantified using root-mean-square (RMS) surface roughness, $R_q$, calculated as the standard deviation of topographic heights.

Following the processing pipeline used for electrostatic and mechanical data, roughness was analyzed in two complementary ways: (1) region-specific trends, where $R_q$ was computed across source, channel, or drain pixels identified by segmentation; and (2) spatial roughness spectra, generated by computing $R_q$ along the slow-scan axis to capture spatial evolution across the device with gate voltage.

This data reveals how surface swelling and roughening co-evolve with electrochemical doping, providing complementary information to mechanical and electrical responses.

*Equivalent circuit model of in-liquid SDM for EGTs*: The local electrostatic force measured by the scanning dielectric microscopy (SDM) probe depends on a combination of geometric, material, and interfacial factors (**Figure S3a**). Key geometric factors are probe shape and tip-sample separation ($z_{lift}$). Relevant material properties include semiconductor film thickness ($h_{sem}$), dielectric constant ($\varepsilon_{sem}$), and conductivity ($\sigma_{sem}$). Interfacial contribution arises from the equivalent tip-electrolyte ($c_{tip}$) and polymer-electrolyte interfacial capacitances ($c_{int}$). Finally, the nature of the underlying substrate, metallic beneath the source/drain and insulating under the channel, primarily determines the directional sensitivity of the measured force. Among these factors, the electrostatic force is largely influenced by the semiconductor conductivity, which can change by orders of magnitude from an insulating to a highly conductive state in response to electrochemical doping.

To provide a transparent picture of these dependencies, we developed a physically grounded equivalent circuit model of in-liquid SDM applied to electrolyte-gated transistors (EGTs) and solved it analytically. **Figure S3b** shows the proposed circuit diagram. A similar approach was first introduced with the original development of in-liquid SDM[48] and has more recently been applied to characterize the electrical properties of confined water.[43]

In the channel region, the bottom insulating substrate prevents field lines from entering perpendicularly, generating a longitudinal electric field within the semiconductor (Figure S3a). We capture this behavior with two perpendicular branches in our equivalent circuit: one representing the local "apex" measurement area and the other non-local microscopic "cone" contribution. These two branches are connected by a longitudinal semiconductor branch. Additional longitudinal elements through the solution or substrate are omitted, as they contribute minimally. Each perpendicular branch terminates in a substrate capacitance. For metallic electrodes (source/drain), these capacitances are effectively infinite, short-circuiting the branches and suppressing lateral current flow through the semiconductor, making the measured electrostatic force sensitive primarily to transversal (*out-of-plane*) conductivity. In contrast, for the channel region on an insulating substrate, the finite substrate capacitance allows the field to penetrate longitudinally, making the electrostatic force mainly sensitive to longitudinal (*in-plane*) transport.

The local electrostatic force acting on the apex of the probe can be obtained as[48]

$$F_{apex} = A_{apex} \frac{1}{2} \frac{\partial c_{apex,sol}}{\partial z} v_{apex,sol}^2 \tag{5}$$

where, $v_{apex,sol}$ is the voltage drop across the apex electrolyte gap (see Figure S3b), $A_{apex}$ is the apex area, and $c_{apex,sol}$ the solution capacitance per unit area under the apex region, given by (assuming a parallel plate model)

$$c_{apex,sol} = \frac{\varepsilon_0 \varepsilon_{sol}}{z} \tag{6}$$

with $\varepsilon_0$ being the vacuum permitivity, $\varepsilon_{sol}$ the electrolyte relative dielectric constant, and $z$ the tip-sample distance. A closed-form analytical expression for $v_{apex,sol}$ can be obtained by solving the equivalent circuit, though the full derivation is omitted due to its cumbersome nature. This solution allows the force-conductivity dependence to be plotted for different device regions (source/drain and channel) and for a wide range of system parameters. **Figure S3c-e** shows the influence of the semiconductor thickness ($h_{sem}$), interfacial capacitance ($c_{int}$), and dielectric constant ($\varepsilon_{sem}$) on the force-conductivity dependence, respectively. In these graphs, the electrostatic force is presented as the capacitance gradient, $dC/dz = 2F_{el}/v_{ac}^2$, where $C$ is the total tip-sample capacitance, $F_{el}$ is the calculated electrostatic force, and $v_{ac}$ is the applied tip-sample voltage. For validation, finite-element numerical calculations were performed following Refs. [33,39] using the same set of parameters (**Figure S3f-h**). Because the simulations assume an infinitely extended lateral domain, the device length in the model was phenomenologically adjusted to a large value to closely match the numerical results. The equivalent circuit model captures the essential features of the simulations while providing a more intuitive framework and a fast way to qualitatively assess trends. Note that the analytical results are plotted on a log-log scale, whereas the numerical simulations are shown on a log-linear scale. This choice reflects the well-established observation that forces calculated for a realistic tip scale approximately logarithmically with those derived under the planar-electrode assumption.[48]

Remarkably, the theoretical trends in electrostatic force closely mirror the experimental behavior seen in OMIEC materials. Increasing the interfacial capacitance, $c_{int}$, enhances the force (Figure S3c), consistent with stronger ionic-electronic coupling in OMIECs, which boosts volumetric capacitance, charge carrier density, and ultimately device transconductance. The influence of the semiconductor thickness, $h_{sem}$, is more nuanced and depends on both device region and conduction state (Figure S3d). Over the source and drain, where the probe primarily senses transverse (*out-of-plane*) transport, increasing $h_{sem}$ lengthens the through-plane path in insulating and semiconducting states, reducing tip-sample coupling and lowering the measured force. In highly conductive films, the vertical voltage drop becomes negligible, so $h_{sem}$ no longer affects the force. This mirrors OMIEC behavior, where thicker films hinder through-plane ionic-electronic coupling at low to mid doping levels but become irrelevant once the film is highly conductive. By contrast, the channel shows an opposite response to $h_{sem}$, reflecting its coupling to lateral rather than vertical transport. Over the channel, increasing $h_{sem}$ in the semiconducting regime enlarges the lateral cross-section for current flow, reducing in-plane impedance and increasing the force. In the insulating limit, there is no lateral conduction, and in the highly conductive state, the in-plane voltage drop is negligible, so $h_{sem}$ has little effect. This aligns with volumetric conduction in OMIECs, where partially doped films benefit from greater volume for charge percolation, thereby sustaining higher on-currents. Finally, the semiconductor dielectric constant, $\varepsilon_{sem}$, primarily affects transverse capacitive coupling (**Figure S3e**). Over source and drain, higher $\varepsilon_{sem}$ increases the displacement current through the material in insulating and semiconducting states, enhancing the electrostatic force. In contrast, over the channel, the measured force is dominated by resistive voltage drops along the semiconductor, so $\varepsilon_{sem}$ has little impact. A higher dielectric constant allows more charge storage across the thickness, improving capacitive coupling and enhancing doping efficiency. Together, these dependencies position electrostatic force as a mechanistic descriptor of OMIECs, linking local transport physics to device-level performance in OECTs.

The characteristic dependence of the electrostatic force on conductivity is sigmoidal (**Figure S3i**) and can be well approximated by the following closed analytical expressions for source/drain and channel regions, respectively

$$F_{apex,S} = A_{apex} \frac{1}{2} \frac{\partial c_{apex,sol}}{\partial z} \left(\frac{c_{apex,eq}}{c_{apex,sol}}\right)^2 \left[\frac{\sigma_{Low,S}^2 + \sigma_{sem,\perp}^2}{\sigma_{High,S}^2 + \sigma_{sem,\perp}^2}\right] v_{ac}^2 \tag{7}$$

$$F_{apex,C} = A_{apex} \frac{1}{2} \frac{\partial c_{apex,sol}}{\partial z} \left(\frac{(A_{apex} + A_{cone})c_{subs}}{A_{apex}(c_{apex,eq} + c_{subs}) + A_{cone}(c_{cone,eq} + c_{subs})}\right) \times \\ \left(\frac{c_{apex,eq}}{c_{apex,sol}}\right)^2 \left[\frac{\sigma_{Low,C}^2 + \sigma_{sem,//}^2}{\sigma_{High,C}^2 + \sigma_{sem,//}^2}\right] v_{ac}^2 \tag{8}$$

where $\sigma_{sem,\perp}$ and $\sigma_{sem,//}$ are the *out-of-plane* and *in-plane* semiconductor conductivities, and we have defined

$$\sigma_{Low,S} = 2\pi f \varepsilon_0 \varepsilon_{sem,\perp} \tag{9}$$

$$\sigma_{High,S} = 2\pi f (c_{apex,eq} + c_{sem,\perp}) h_{sem} \tag{10}$$

and

$$\sigma_{Low,C} = 2\pi f \left[ \frac{A_{apex} A_{cone}}{(A_{apex} + A_{cone})} (c_{cone,eq} + c_{subs}) \right] \frac{L_{sem}}{h_{sem} W_{sem}} \tag{11}$$

$$\sigma_{High,C} = 2\pi f \left[ \frac{A_{apex} A_{cone} (c_{apex,eq} + c_{subs})(c_{cone,eq} + c_{subs})}{A_{apex}(c_{apex,eq} + c_{subs}) + A_{cone}(c_{cone,eq} + c_{subs})} \right] \frac{L_{sem}}{h_{sem} W_{sem}} \tag{12}$$

with

$$c_{sem,\perp} = \frac{\varepsilon_0 \varepsilon_{sem,\perp}}{h_{sem}} \tag{13}$$

$$c_{subs} = \frac{\varepsilon_0 \varepsilon_{subs}}{h_{subs}} \tag{14}$$

$$c_{cone,sol} = \frac{\varepsilon_0 \varepsilon_{sol}}{H_{cone} + z} \tag{15}$$

$$c_{apex,eq} = \left( \frac{1}{c_{apex,sol}} + \frac{1}{c_{tip}} + \frac{1}{c_{int}} \right)^{-1} \tag{16}$$

$$c_{cone,eq} = \left( \frac{1}{c_{cone,sol}} + \frac{1}{c_{tip}} + \frac{1}{c_{int}} \right)^{-1} \tag{17}$$

Here, $f$ is the frequency of applied AC tip voltage; $A_{cone}$ the cone equivalent area; $\varepsilon_{sem,\perp}$ and $\varepsilon_{sem,//}$ the *out-of-plane* and *in-plane* semiconductor relative dielectric constants, respectively; $W_{sem}$ and $L_{sem}$ the semiconductor width and (characteristic) length, respectively; $c_{tip}$ the equivalent tip/electrolyte interfacial capacitance; $\varepsilon_{subs}$ and $h_{subs}$ are the substrate relative dielectric constant and thickness, respectively; and $H_{cone}$ the cone height. Equations (9)-(12) give the characteristic conductivities (strictly speaking, the first turnover conductivity is at $\sigma_{Low,S}/2$, and the second one at $2\sigma_{High,S}$). The agreement between the approximate equations (7) and (8) and the exact equivalent circuit solution is remarkable (see the black dashed lines in Figure S3i).

The analysis reveals three distinct regimes in both source/drain and channel regions, separated by two characteristic conductivities in the transverse and longitudinal directions, respectively (Figure S3i). Over the source and drain, when the *out-of-plane* conductivity is low ($\sigma_{sem,\perp} < \sigma_{Low,S}$; material's dielectric relaxation threshold), the material behaves as a dielectric (insulating regime). At intermediate values ($\sigma_{Low,S} < \sigma_{sem,\perp} < \sigma_{High,S}$), it acts as a lossy dielectric (semiconducting regime), and at high conductivities ($\sigma_{sem,\perp} > \sigma_{High,S}$; Maxwell-Wagner transition), the material appears "metallic" to the SDM probe, with negligible voltage drop across it (conducting regime). Over the channel, these regimes are governed by the *in-plane* conductivity: low conductivity ($\sigma_{sem,//} < \sigma_{Low,C}$) produces dielectric behavior along the channel length (while remaining vertically conductive), intermediate values ($\sigma_{Low,C} < \sigma_{sem,//} < \sigma_{High,C}$) correspond to lossy semiconducting behavior, and high conductivity ($\sigma_{sem,//} > \sigma_{High,C}$) renders the channel effectively "metallic" with no appreciable longitudinal voltage drop. Together, these regimes define the sigmoidal dependence of electrostatic force on conductivity and establish the physical basis for interpreting SDM measurements in EGTs.

*Organic electrochemical transistors fabrication*: BBL-based OECTs were fabricated following a previously reported protocol,[49] which uses a sacrificial parylene-C layer to define the device channel. Standard microscope glass slides were cleaned via sequential sonication in acetone, deionized water, and isopropyl alcohol, followed by drying under nitrogen flow. Gold source and drain electrodes (5 nm Cr / 50 nm Au) were thermally evaporated and patterned by photolithography and wet etching. A 4 μm-thick layer of parylene C (PaC), deposited together with a small amount of 3-(trimethoxysilyl)propyl methacrylate (A-174 Silane) for enhanced adhesion, served as an insulating base layer, suppressing parasitic capacitive effects at the metal-liquid interface. To enable channel patterning, a sacrificial PaC bilayer was prepared: first, a 2% solution of Decon-90 surfactant was spin-coated as an anti-adhesive, followed by deposition of a second 4 μm-thick PaC layer. A 10 μm-thick positive photoresist (AZ10XT520CP) was spin-coated onto the parylene stack to protect it during plasma etching. Photolithography was then used to define the channel and contact pad regions. After developing the exposed photoresist, reactive ion etching (150 W, $O_2$: 500 sccm, $CF_4$: 1000 sccm, 510 s) was used to etch through both parylene layers and the photoresist, exposing the gold electrodes and defining wells for the active material, while leaving the rest of the surface passivated. Channels with two aspect ratios were fabricated: W/L = 20 μm/10 μm and 100 μm/10 μm. The BBL solution in methane sulfonic acid (MSA) was spin-coated over the entire substrate to form a ~25 nm-thick semiconductor film. The sacrificial parylene layer was then peeled off, removing excess polymer and leaving patterned BBL films confined to the channel wells between source and drain electrodes.

**Data availability**

The data that support the findings of this study will be made openly available in Zenodo at DOI:10.5281/zenodo.16890860 upon publication.

**Acknowledgements**

This work received funding from the European Union's Horizon 2020 research and innovation program under the Marie Skłodowska-Curie grant agreement No 813863 (BORGES), the EIC Pathfinder PRINGLE project (grant agreement No 101046719), from the Spanish Ministerio de Economıa, Industria y Competitividad, and EU FEDER, through grant no. PID2019-110210GB-I00 (BIGDATASPM), from the Ministerio de Ciencia e Innovación through grant no. PID2022-142297NB-I00 (BIOMEDSPM40), from the Generalitat de Catalunya through CERCA, and from the ICREA foundation (ICREA Academia award to G.G.). S.T. acknowledges support from Joerg Barner (JPK) for automating AFM operations. H.-Y.W., C.-Y.Y., and S.F. acknowledge financial support from the Knut and Alice Wallenberg Foundation (2021.0058) and the Swedish Research Council (2022-04053, 2022-04553, and 2024-04871).

**Author Contributions**

S.T. led the investigation, designed and developed the automated multimodal experimental setup, performed all the measurements and data analysis, interpreted the results, drafted and revised the manuscript. H.-Y.W., C.-Y.Y., and S.F. designed and fabricated the BBL OECTs. R.M.-S. carried out numerical simulations. S.F. contributed ideas, proposed experiments, and edited the manuscript. A.K. led the collaboration. A.K. and G.G. supervised the investigation, managed the project, and edited the manuscript. G.G. developed the equivalent circuit model of in-liquid SDM. All authors contributed to discussions and provided feedback on the final manuscript.

**Conflict of Interest**

The authors declare no conflict of interest.

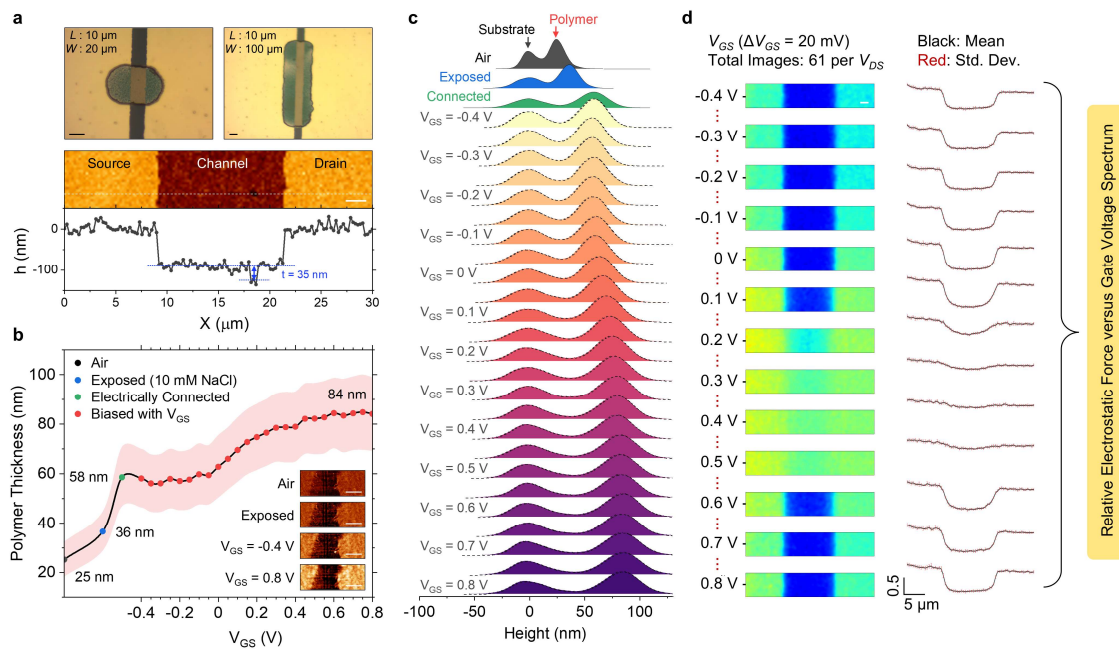

**Figure S1. Device architecture, polymer electrochemical swelling behaviour, and multimodal spectral reconstruction process. a**, Optical and atomic force microscopy (AFM) images of BBL OECTs with fixed channel length ($L = 10$ μm) and two channel widths ($W = 20$ μm, left; 100 μm, right). The AFM topography image (middle, $30 \times 5.625$ μm$^2$, $128 \times 24$ pixels) was acquired in 10 mM NaCl solution prior to multimodal mapping (Figures 1 and 2, main text). A height profile (bottom) through a hole in the polymer film reveals the film thickness ($t = 35$ nm). **b**, Polymer thickness under dry and hydrated states, showing both passive and active swelling. The dry film (~25 nm) swells passively to ~36 nm when immersed in electrolyte, increases to ~58 nm after establishing electrical connections, and expands further up to ~84 nm during electrochemical gating (active swelling). Thickness is determined by scratching the film above the source electrode to expose the gold substrate. The shaded error band indicates surface roughness (standard deviation of polymer-region height). Selected AFM images are shown in the inset. **c**, Histogram of pixel height values from AFM images in **b**, with peaks corresponding to the gold substrate (zero height) and the polymer film. Peak position yields thickness; width reflects surface roughness. **d**, Workflow for reconstructing gate-dependent relative electrostatic force spectra. Average line profiles from sequential electrical images at fixed gate voltages are compiled into a 2D spectral map. For clarity, only a subset of the 61 images (per $V_{DS}$) is shown, corresponding to Figure 1d in the main text. The same procedure is applied to mechanical indentation slope and surface roughness datasets. Scale bars: optical images (black, 10 μm), AFM images (white, 2 μm).

Page - 36 - of 45

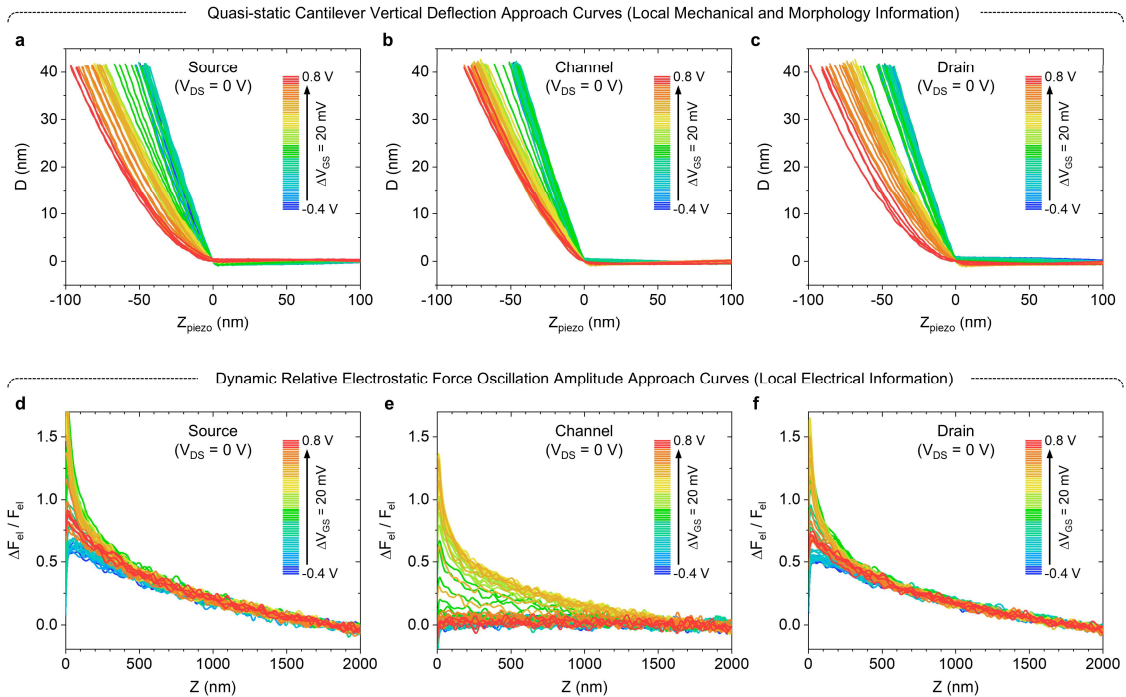

**Figure S2. In-liquid multimodal scanning dielectric microscopy (SDM) approach curves at representative device locations.** Representative quasi-static and dynamic approach curves acquired in the source, channel, and drain regions of an operating OECT under aqueous electrolyte. **a-c**, Quasi-static cantilever vertical deflection ($D$) versus piezo displacement ($Z_{piezo}$) during tip approach at the center of the source (**a**), channel (**b**), and drain (**c**) regions. Curves are shown for gate voltages ($V_{GS}$) from -0.4 V to +0.8 V in 20 mV steps ($\Delta V_{GS}$) at zero drain bias ($V_{DS}$ = 0 V). **d-f**, Corresponding dynamic relative electrostatic force amplitude ($\Delta F_{el}/F_{el}$ with $Z_{ref}$ = 1800 nm) versus tip-sample distance ($Z$), obtained from the oscillatory component of the cantilever response at the same positions as in **a-c**. The quasi-static curves reflect local mechanical deformation and morphology, whereas the dynamic curves capture the gate-dependent local electrical response. These representative point spectra illustrate the signal origin of the multimodal SDM probe, where analogous curves were acquired at every pixel to reconstruct spatially resolved maps of electrostatic, mechanical, and topographic properties at each operating bias. Acquisition parameters: setpoint = 50 nN; Z-length = 3000 nm; pixel time = 84 ms, comprising extend (75 ms), retract (4 ms), and miscellaneous (5 ms) segments. Other parameters are as in Figure 1 of the main text.

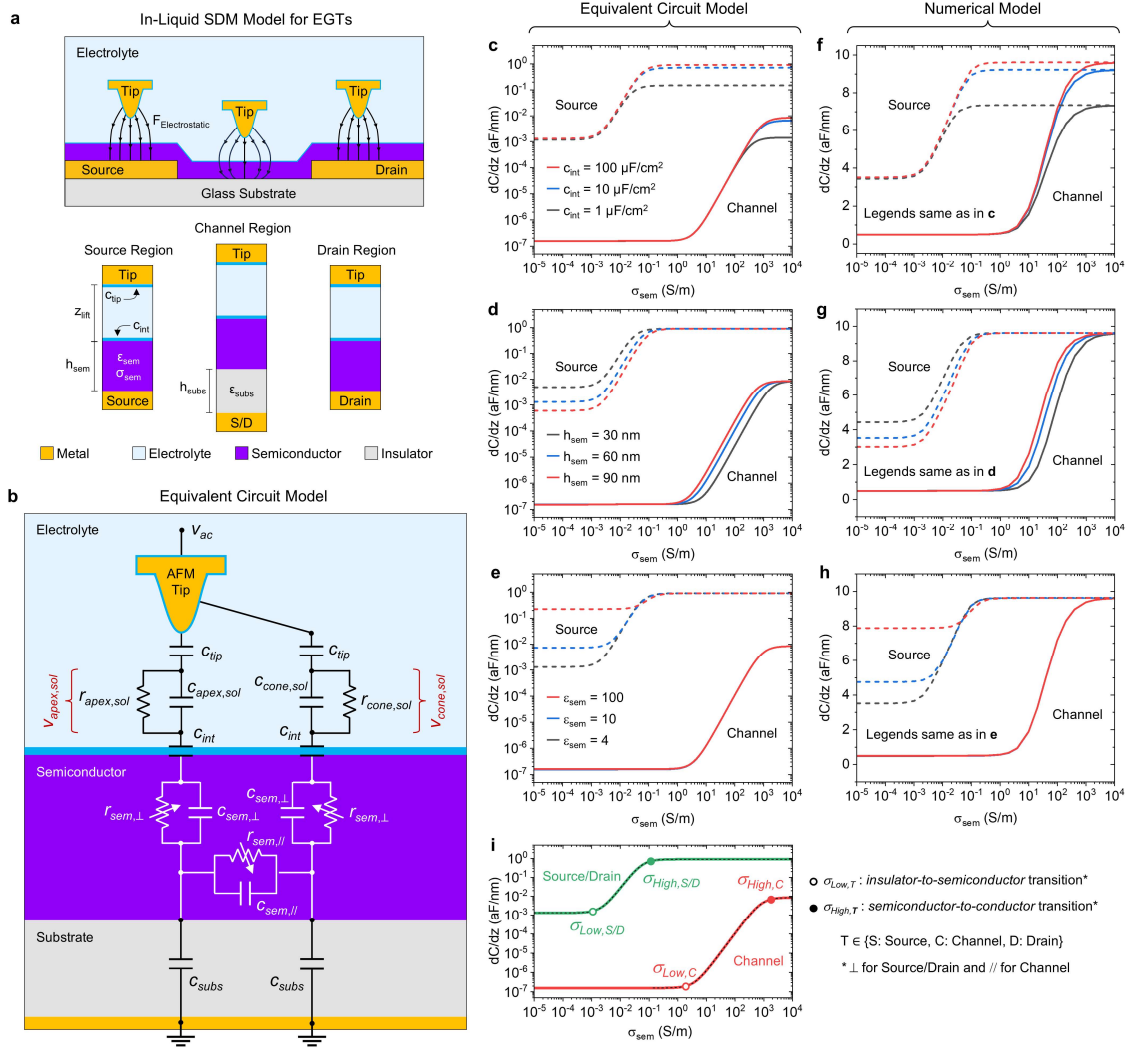

**Figure S3. Modelling framework linking nanoscale electrostatic force response to material and device parameters. a**, Conceptual illustration of the electrostatic field distribution and vertical device stack encountered by the scanning tip over the source, channel, and drain regions. **b**, Physically-grounded equivalent circuit model that captures the coupling between the tip, electrolyte, and semiconductor, developed to provide an intuitive understanding of how system parameters influence the measured electrostatic force, insight not readily accessible from numerical models. **c-e**, The evolution of apex (local) electrostatic force with semiconductor conductivity ($\sigma_{sem}$) obtained from equivalent circuit model for varying effective interfacial capacitance (**c**, $c_{int}$), semiconductor thickness (**d**, $h_{sem}$), and semiconductor dielectric constant (**e**, $\varepsilon_{sem}$). Other nominal model parameters defined in the *Methods* section are: $c_{tip}$ = 2.7 μF/cm², $c_{int}$ = 100 μF/cm², $z_{lift}$ = 20 nm, $h_{sem}$ = 60 nm, $W_{sem}$ = 100 μm, $L_{sem}$ = 10 μm × 10⁵ (phenomenologically adjusted), $h_{subs}$ = 10 μm, $\varepsilon_{sem}$ = 4, $\varepsilon_{subs}$ = 7, $\varepsilon_{sol}$ = 78, $\sigma_{sol}$ = 1 μS/m, $A_{apex}$ = 2·π·R² with R = 30 nm, $A_{cone}$ = π·H²·tan(θ)/cos(θ) with H = 12 μm and θ = 25°,

and $f$ = 10 MHz. **f-h**, Results obtained from finite element numerical simulations for an axially symmetric laterally infinite thin-film, validating the simplified equivalent circuit model. **i**, Definition of characteristic conductivity points marking key transitions in the electrostatic force versus conductivity response for the source/drain (green) and channel (red) regions. Dashed lines represent fits using equations (7) and (8) in the *Methods* section, yielding $\sigma_T$ points for source/drain ($\sigma_{Low,S/D}$ = 1.11 × 10$^{-3}$ S/m; $\sigma_{High,S/D}$ = 0.12 S/m) and channel ($\sigma_{Low,C}$ = 1.88 S/m; $\sigma_{High,C}$ = 1758.12 S/m) regions.

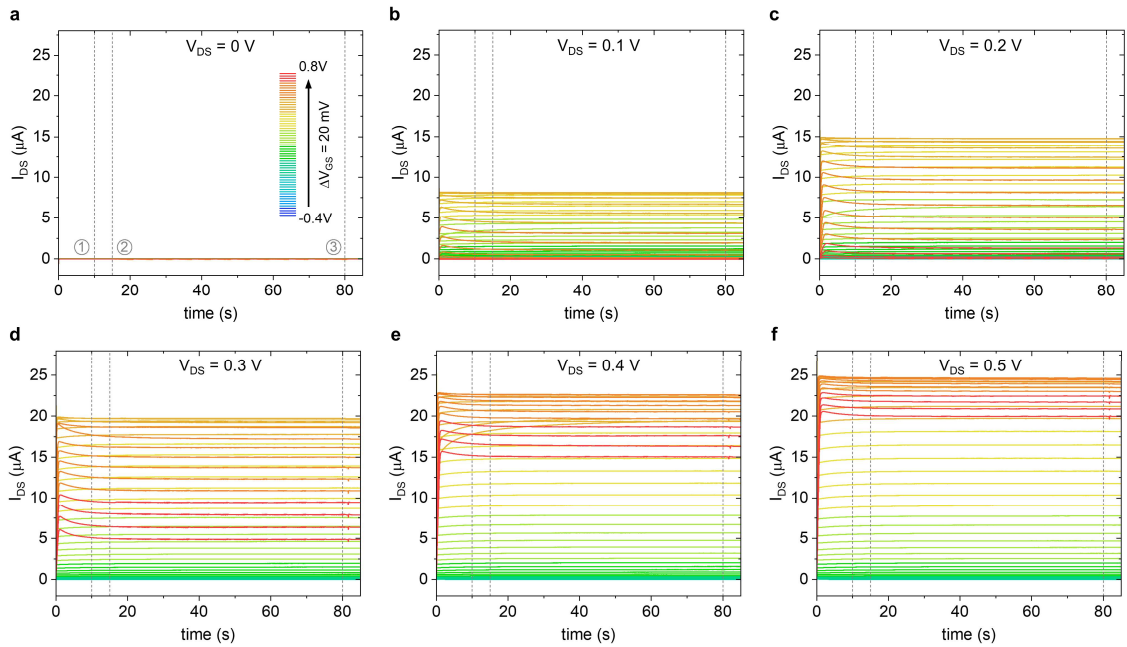

**Figure S4. Drain current evolution during operando in-liquid SDM imaging.** Time evolution of drain current ($I_{DS}$) recorded during multimodal SDM measurements at gate voltages ($V_{GS}$) from -0.4 V to +0.8 V in steps of 20 mV ($\Delta V_{GS}$), for drain voltages ($V_{DS}$) of 0 V (**a**), 0.1 V (**b**), 0.2 V (**c**), 0.3 V (**d**), 0.4 V (**e**), and 0.5 V (**f**). Legends are common to all panels and shown in **a**. At $t = 0$ s, fixed $V_{GS}$ and $V_{DS}$ biases are applied; after 10 s (dotted line 1), the SDM probe approaches the device surface. Imaging begins following a 5 s wait (dotted line 2), during which multimodal approach curves are acquired that last 65 s. The probe is then withdrawn (dotted line 3), and after a 5s delay, all bias voltages are turned off. Each trace captures the global device response of the underlying coupled electronic-ionic dynamics at specific biasing conditions, providing the basis for reconstructing transfer characteristics that can be correlated with simultaneously acquired spatially resolved physical property maps. Data shown here correspond to measurements in Figures 1 and 2 of the main text and in Figure S5.

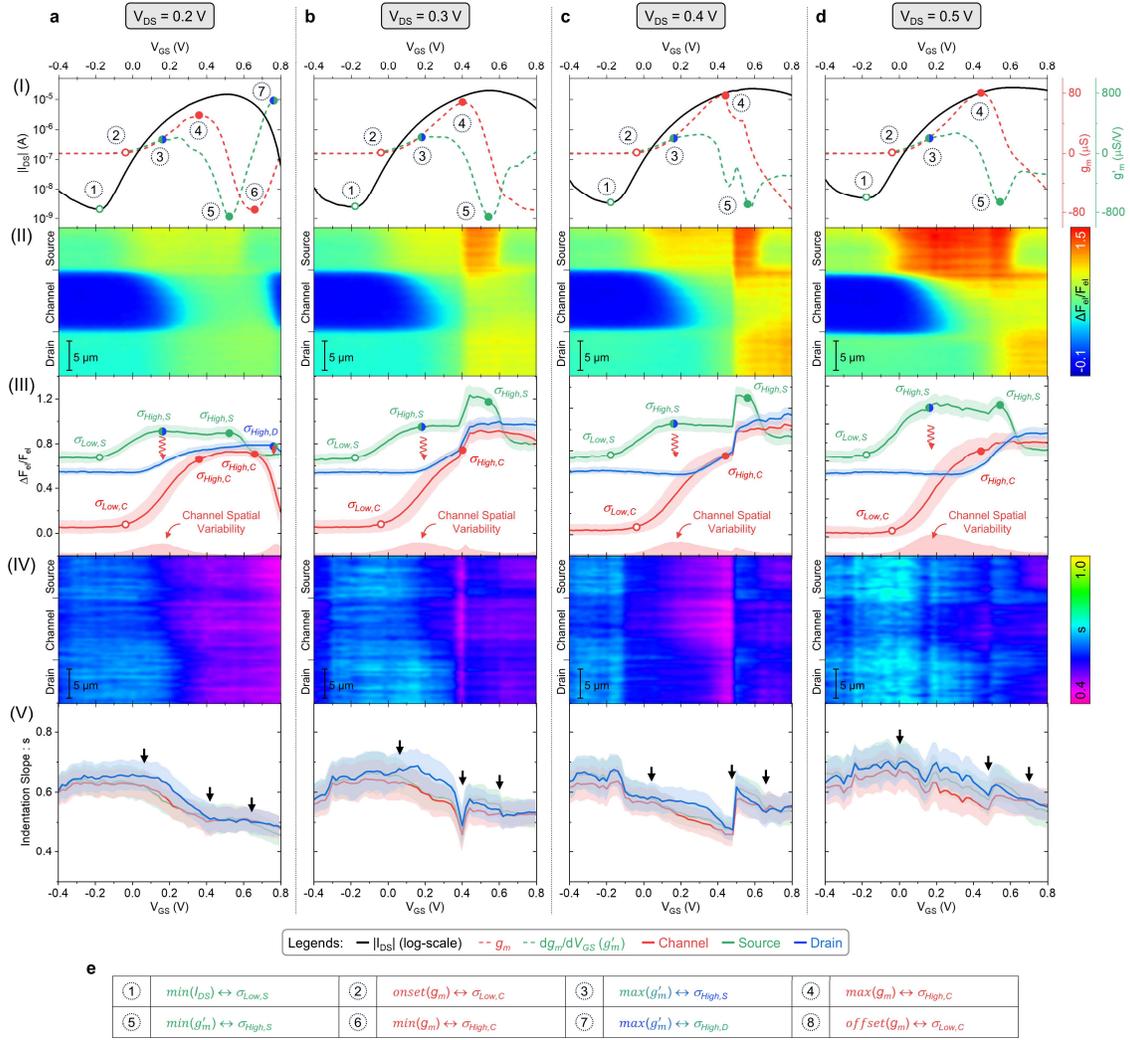

**Figure S5. Multimodal mapping of local device dynamics across varying drain voltages.** Local measurements are correlated with current-voltage transfer curves at $V_{DS}$ = 0.2 V (**a**), 0.3 V (**b**), 0.4 V (**c**), and 0.5 V (**d**). Each dataset contains multiple panels showing: (I) the reconstructed transfer curve (black) with first (transconductance $g_m$, red) and second ($dg_m/dV_{GS}$ or $g'_m$, green) derivatives; (II) spatially resolved gate-dependent electrostatic force spectra; (III) region-averaged trends of electrostatic force for source (green), channel (red), and drain (blue); (IV) spatially resolved gate-dependent mechanical spectra; and (V) region-averaged trends of indentation slope. Key operating points identified in the transfer curves (panel I) correspond to characteristic local conductivity states across device regions (panel III), and significant mechanical shifts are indicated by black arrows in panel V. **e**, summarizes the correspondence between global transfer curve features and local physical states. These datasets complement Figure 2 in the main text.

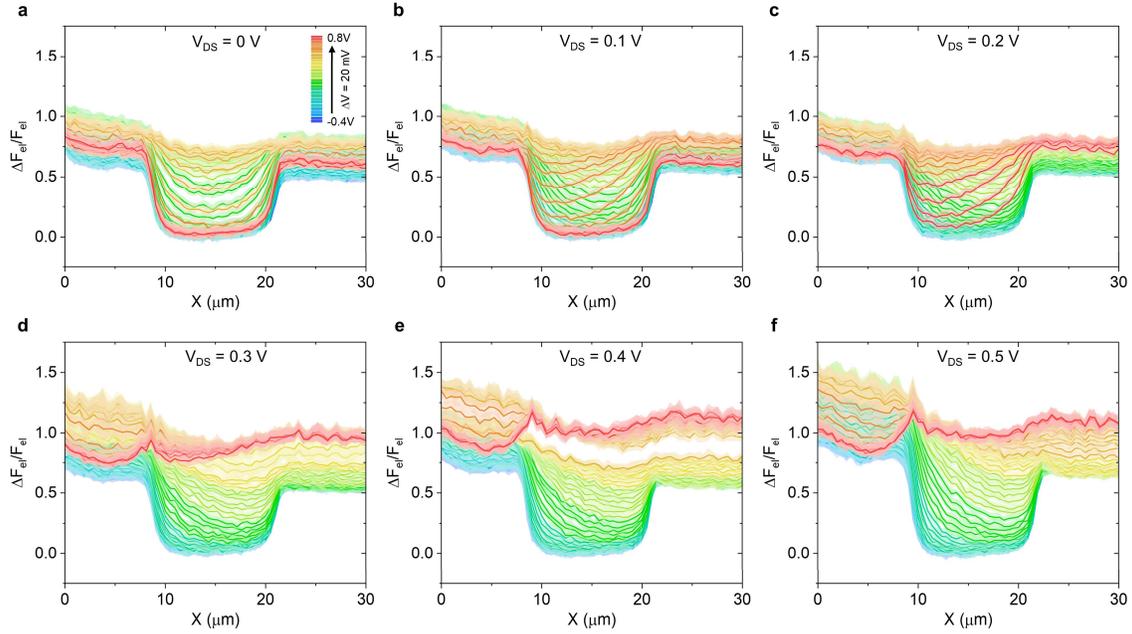

**Figure S6. Spatially resolved electrostatic force defining the internal state of an operating OECT.** Average profiles of the relative electrostatic force along the source-channel-drain axis at increasing drain bias: $V_{DS}$ = 0 V (**a**), 0.1 V (**b**), 0.2 V (**c**), 0.3 V (**d**), 0.4 V (**e**), and 0.5 V (**f**), corresponding to the measurements shown in Figures 1 and 2 of the main text and Figure S5. Curves correspond to different gate voltages ($V_{GS}$, colour-coded, from -0.4 V to 0.8 V in 20 mV steps). Each profile represents the average along the slow scan axis extracted from 2D spatial maps, while the shaded region indicates the standard deviation, reflecting lateral heterogeneity across the device width. The pronounced bias-dependent spatial modulation of the local electrical properties across source, channel, and drain defines the *internal state* of the OECT at a given operating point. With increasing $V_{DS}$, the profiles become progressively asymmetric, indicating a redistribution of local electrical properties along the channel. Correlating these complex local source-channel-drain dynamics with simultaneously acquired global device response enables identification of how spatially distributed *internal states* are embedded in the device transfer characteristics.

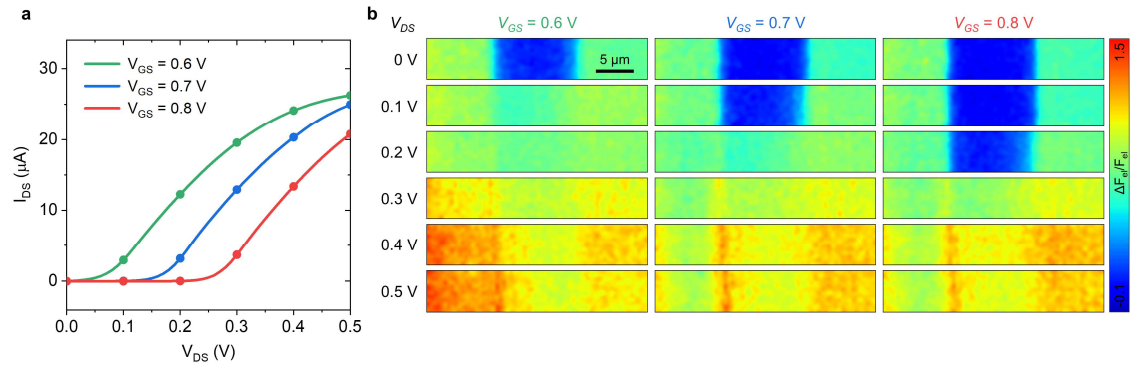

**Figure S7. Spatially resolved origin of gate-dependent thresholds in high-doping output curves.** Output characteristics ($I_{DS}$ versus $V_{DS}$) of BBL OECTs in the negative transconductance or high doping regime (**a**) show that higher gate voltages require increasingly larger drain voltages to turn on the device. Local electrostatic force maps (**b**) reveal the spatial origin of this behavior: at high gate bias, the channel enters an insulating state due to reduced carrier mobility (blue coloured region in **b**). The applied drain voltage compensates for the local effective gate potential, reactivating conduction in the channel. Regions near the source experience the highest effective local gate potential and remain less conductive; therefore, larger drain voltages are required to achieve compensation relative to other regions.

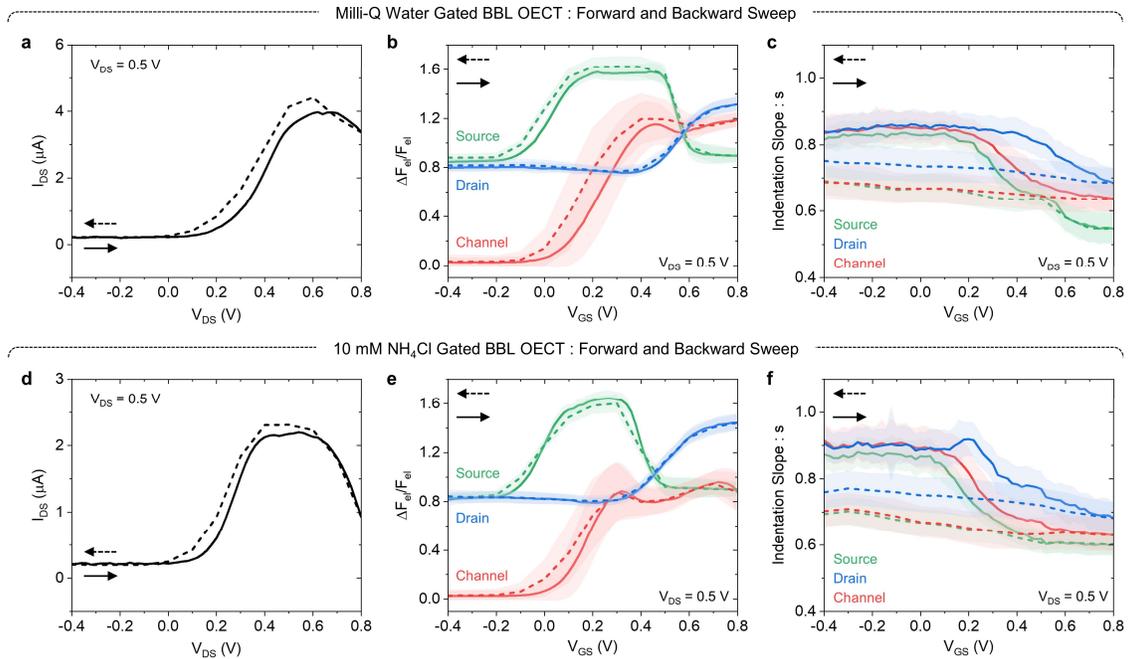

**Figure S8. Hysteresis in gate-voltage sweeps arises from persistent yet partially reversible mechanical changes.** Transfer curves and local property evolution of BBL OECTs during forward and backward gate-voltage sweeps. **a-c**, Devices gated with Milli-Q water: drain current versus gate voltage transfer curve (**a**) and corresponding local electrical (**b**) and mechanical (**c**) responses. Additional mechanical changes over the source at high gate voltages are reversible in the range $0.5\ V \leq V_{GS} \leq 0.8\ V$ (see green curve in **c**). **d-e**, Same measurements as in **a-c** for the device gated with 10 mM NH$_4$Cl, showing ion-specific effects with no additional mechanical change observed at high gate voltages over source.

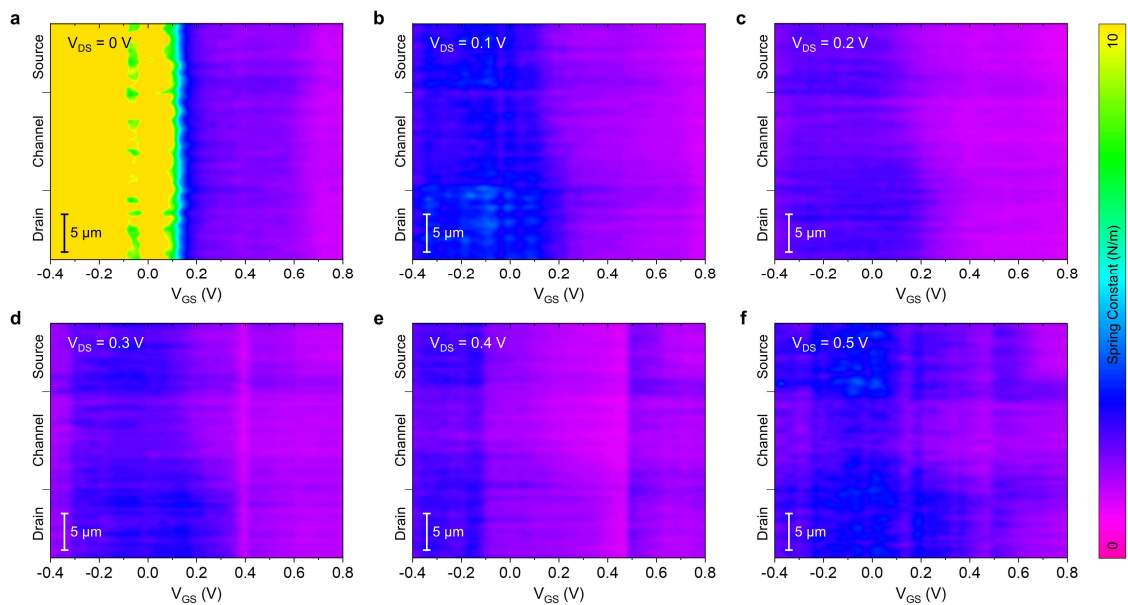

**Figure S9. Gate and drain bias-dependent effective spring constant of BBL OECTs. a-f**, Effective spring constant of the OECT device starting from a pristine condition and following gate-voltage cycling during multimodal measurements, shown for increasing drain biases: $V_{DS}$ = 0 V (**a**), 0.1 V (**b**), 0.2 V (**c**), 0.3 V (**d**), 0.4 V (**e**), and 0.5 V (**f**). Values are derived from the indentation slope datasets corresponding to Figures 1 and 2 of the main text and Figure S5, capturing the local mechanical response across the device.